# Chemical Interaction and Electronic Structure in a Compositionally Complex Alloy: a Case Study by means of X-ray Absorption and X-ray Photoelectron Spectroscopy

S. Kasatikov*[a,b], A. Fantin[b,c], A. M. Manzoni[b,d], S. Sakhonenkov[a], A. Makarova[e], D. Smirnov[f], E. O. Filatova[a] and G. Schumacher[b,c]

[a] Institute of Physics, St. Petersburg State University, 198504 St. Petersburg, Russia

[b] Helmholtz-Zentrum Berlin, D-14109 Berlin, Germany

[c] Technische Universität Berlin, D-10623 Berlin, Germany

[d] Bundesanstalt für Materialforschung und -prüfung, Abteilung Werkstofftechnik, 12205 Berlin, Germany

[e] Institut für Festkörper- und Materialphysik, Technische Universität Dresden, 01062 Dresden, Germany

[f] Physikalische Chemie, Institut für Chemie und Biochemie, Freie Universität Berlin, 14195 Berlin, Germany

*Corresponding author: S. Kasatikov, st031736@student.spbu.ru

**Abstract**

Chemical interaction and changes in local electronic structure of Cr, Fe, Co, Ni and Cu transition metals (TMs) upon formation of an $Al_8Co_{17}Cr_{17}Cu_8Fe_{17}Ni_{33}$ compositionally complex alloy (CCA) have been studied by X-ray absorption spectroscopy and X-ray photoelectron spectroscopy. It was found that upon CCA formation, occupancy of the Cr, Co and Ni 3d states changes and the maximum of the occupied and empty Ni 3d states density shifts away from Fermi level ($E_f$) by 0.5 and 0.6 eV, respectively, whereas the Cr 3d empty states maximum shifts towards $E_f$ by 0.3 eV, compared to the corresponding pure metals. The absence of significant charge transfer between the elements was established, pointing to the balancing of the 3d states occupancy change by involvement of delocalized 4s and 4p states into the charge redistribution. Despite the expected formation of strong Al–TMs covalent bonds, the Al role in the transformation of the TMs 3d electronic states is negligible. The work demonstrates a decisive role of Cr in the Ni local electronic structure transformation and suggests formation of directional Ni–Cr bonds with covalent character. These findings can be helpful for tuning deformation properties and phase stability of the CCA.

**Keywords**

High-Entropy Alloys; Compositionally Complex Alloys; X-ray Spectroscopy; Electronic Structure; Chemical Bonding

**Introduction**

During the last 15 years, a new emerging family of metallic alloys, i.e. high entropy alloys (HEAs) and compositionally complex alloys (CCAs), revealed outstanding mechanical properties, thus great potential for many applications opening a new door for modern alloy design.[1-4] Knowledge of electronic structure and the role of each element in the HEA/CCA super-saturated solid solution structure are a basic prerequisite for the understanding of such unique physical, chemical, mechanical and thermodynamic properties.

In the review by Miracle & Senkov,[1] the authors put emphasis on the fact that two of four "core effects" accountable for the HEAs' and CCAs' particularity i) the lattice distortion and ii) the 'cocktail' effect are interconnected with electronic structure. The lattice distortion effect is claimed to be strongly interrelated with local chemical ordering, dependent on chemical interaction between the alloyed elements.[5-7] The 'cocktail' effect is considered as a complex synergetic mechanism that is responsible for outstanding physical, functional and structural properties of a vast set of multi-principal elements.[1]

In the recent work by George et al. it is highlighted that exploration of local chemical ordering in HEAs is crucial for advancing in the field.[3] It is expected that local chemical ordering affects stacking-fault energy and dislocation mobility and ,therefore, is decisive in controlling mechanical properties.[4,8] The complexity of the phenomenon requires deep understanding of elements interaction at both atomic and electronic levels.

However, the substantial lack of experimental and theoretical data appears to be a large obstacle to further development of the multi-principal element alloy design strategy, because of the tremendous amount of possible HEA/CCA compositions. For the same reason, answering the question about the individual roles of each element in HEA/CCA formation is of primary importance. Thus, the development of fundamental approaches based on revealing chemical roles of alloying elements, especially their role in electronic structure, are imperative.

Recent studies have already drawn attention to the prediction of HEA/CCA mechanical properties on the basis of electronic structure properties of alloying elements, such as the valence electron concentration (VEC) parameter.[9,10] Chen et al. proposed and verified a method based upon the relationship between the VEC parameter and phase structure to balance ductility and strength of HEAs using examples of two systems: $(CoCrCuFeNi)_{100-x}Mo_x$ and $(AlCoCrFeNi)_{100-x}Ni_x$.[10] The authors demonstrated the important role of the VEC parameter for HEAs: matrix strength is improved by adding an element with a VEC lower than the average VEC of the matrix (with *bcc* fracture increase), while ductility is improved by adding an element with a higher VEC than the average VEC of the matrix (with *fcc* fracture decrease).

Opposite results were obtained by Sheikh et al. for a *bcc* refractory HEA (RHEA) using an electron theory: "Intrinsically ductile RHEAs can be developed by alloying elements from group VI or group V, with elements from group V or group IV, or in other words by decreasing the number of valence electrons (s + d electrons), in single-phase *bcc* solid solutions".[9] To rationalize the obtained correlation, the authors exploited the concepts proposed previously by Chan and Qi & Chrzan for refractory binary alloys.[11,12]

Chan claimed that decreasing VEC in Nb-based alloys can reduce the Peierls-Nabarro barrier energy and hence improves the tensile ductility by increasing the dislocation mobility.[11] Similarly to Chan, Qi & Chrzan theoretically demonstrated a ductilization of W- and Mo-based *bcc* alloys by decreasing their VEC.[12] The phenomenon is explained by a shift of the Fermi level ($E_f$) relative to the band structure, facilitating shear instability and hence inducing brittle-to-ductile transition. The authors also mentioned possible involvement of Jahn-Teller distortion that assists shear instability and is promoted by the change in valence band (VB) occupancy and increase in density of states at $E_f$.

More literature about the VEC influence on HEA and other alloys phase stability and their mechanical behavior can be found elsewhere.[13-16]

Nevertheless, although the concept of VEC adjustment for tailoring HEA properties seems to be promising, the starting assumption of using tabulated VEC parameters for pure elements is quite simplistic. This approach does not allow for possible valence state density and electron charge redistribution upon alloying due to chemical interaction between the elements.

Indeed, there are many evidences of valence state redistribution upon binary and ternary metallic alloy formation due to chemical interaction of the elements.[13,17-29] Modification of the density of states (DOS) shape in the vicinity of $E_f$ was reported to correlate with changes in material properties such as ductility/brittleness, elastic moduli,[20,26,28] micro-hardness,[26] grain boundary cohesive strength,[20] energy of martensitic transformation,[28] heat formation, cohesive energy and phase stability.[13,19,21,22]

DOS transformation upon alloy formation is mainly attributed to hybridization between valence electronic states of alloying elements. In turn, the hybridization points to formation of directional bonds with covalent character, decisive for the deformational properties of alloys.[20-22,26,28,30-32] It was demonstrated that variation of bonding character from metallic to covalent in an alloy, by decreasing the concentration of a particular element involved in the bonding, allows tailoring of the material mechanical properties.[20,26,28,29,32]

The present work is an experimental study on the electronic structure and chemical bonding in a CCA, focused on chemical ordering of alloying elements, bonding character between a particular pair of elements, and influence of a specific element on the valence electron states distribution in the CCA. Deeper understanding of such aspects is expected to facilitate the selection of promising compositions of medium-entropy alloys (MEAs), HEAs and CCAs.

In the current study we attempt to reveal these fundamentals behind HEA and CCA formation by focusing on an *fcc* structured $Al_8Co_{17}Cr_{17}Cu_8Fe_{17}Ni_{33}$ CCA. $Al_8Co_{17}Cr_{17}Cu_8Fe_{17}Ni_{33}$ has been selected because i) it shows a simple *fcc* structure, and ii) it is single-phase at a large temperature window (about 900 – 1250°C), segregating in a two-phase system only below 900°C.[33,34] In this study, only the single phase domain is addressed.

The alloy is studied along with its corresponding pure transition metal (TM) elements, binary ($Ni_{90}Al_{10}$), ternary (CrFeNi) and quinary ($Co_{19}Cr_{19}Cu_9Fe_{19}Ni_{34}$, $Al_{9.6}Co_{20.4}Cu_{9.6}Fe_{20.4}Ni_{40}$) single-phase sub-alloys by means of near-edge X-ray absorption fine structure (NEXAFS) and X-ray photoelectron (XPS) spectroscopy. A chemical bonding and electronic structure study of $Al_8Co_{17}Cr_{17}Cu_8Fe_{17}Ni_{33}$ is of high relevance due to the reported chemical ordering in the same,[34] and in a similar alloy, $Al_{1.3}CoCrCuFeNi$,[5,6] and for the well-known strong chemical interaction between Al and 3d TMs.[18-26,35]

**Experimental**

Al, Co, Cr, Cu, Fe and Ni elements of high purity (99.99 %) were melted together in appropriate ratios in a high frequency vacuum induction furnace. Four alloys have been prepared for this study and their preparation parameters for obtaining homogeneity and single-phase state are listed in Table 1. The obtained ingots were then cut into slices of 1.5 mm thickness and homogenized and quenched as described in Table 1.

Table 1. Alloys investigated in this study, and their corresponding homogenization parameters and notations used.

| Alloy (at. %) | Homogenization | Notation |
|---|---|---|
| $Al_8Co_{17}Cr_{17}Cu_8Fe_{17}Ni_{33}$ | 1250°C 1 h – water quenching | CCA |
| $Co_{19}Cr_{19}Cu_9Fe_{19}Ni_{34}$ | 1250°C 1 h – water quenching | $CCA_{sans}Al$ |
| $Al_{9.6}Co_{20.4}Cu_{9.6}Fe_{20.4}Ni_{40}$ | 1250°C 1 h – water quenching | $CCA_{sans}Cr$ |
| CrFeNi | 1100°C 10 h – water quenching | CrFeNi |
| $Ni_{90}Al_{10}$ | 1250°C 10 h – water quenching | $Ni_{90}Al_{10}$ |

Optical microscopy, scanning electron microscopy, transmission electron microscopy, atom probe tomography and conventional X-ray diffraction were used to prove the homogeneity of the CCA at different length scales. Results are summarized and described elsewhere.[34]

Prior to the NEXAFS measurements, the specimens were pre-characterized by XRD with a Bruker D8 Advance instrument in Bragg-Brentano geometry, equipped with a LYNXEYE detector and a nickel filter (0.5 μm). The characteristic radiation lines used were Cu $K\alpha_1$ (1.5406 Å) and Cu $K\alpha_2$ (1.5444 Å). Phase identification was carried out with the ICDD PDF2 database in the *EVA14* software.[36] Structural refinements were carried out using the software *TOPAS*.[37]

The NEXAFS and XPS measurements were performed at the RGL-PES station of the Russian-German beam-line (RGBL) at the synchrotron light source BESSY II of Helmholtz-Zentrum Berlin.[38] Prior to the NEXAFS and XPS measurements the surface of each specimen was cleaned *in situ* via sputtering with 1.5 kV Ar$^+$ ions at the 30° grazing incident angle of the ions. Additionally, after the sputtering the reference pure metals (Cr, Fe, Co, Ni and Cu) were heat treated *in situ* for 20 min. at 700 °C. The NEXAFS and XPS spectra were measured at an incident photon angle of 55° with energy resolution better than E/ΔE = 2000. The NEXAFS spectra were obtained in two acquisition modes: total electron yield (TEY) and fluorescence yield (FY). Calibration of the photon energy scale was performed by measuring Au $4f_{7/2}$ photoelectron line (84.0 eV) of a gold single crystal.

Normalization of the spectra was performed taking into account the synchrotron ring current recorded simultaneously with the spectra and the photon-flux of the incident beam obtained by measuring absorption of a gold single crystal in the TEY mode.

**Results**

**1. Study of the structure and homogeneity**

In Figure 1, the XRD patterns of the specimens measured in this work – without taking into account pure metal references – are presented as a function of the scattering angle *2θ*.

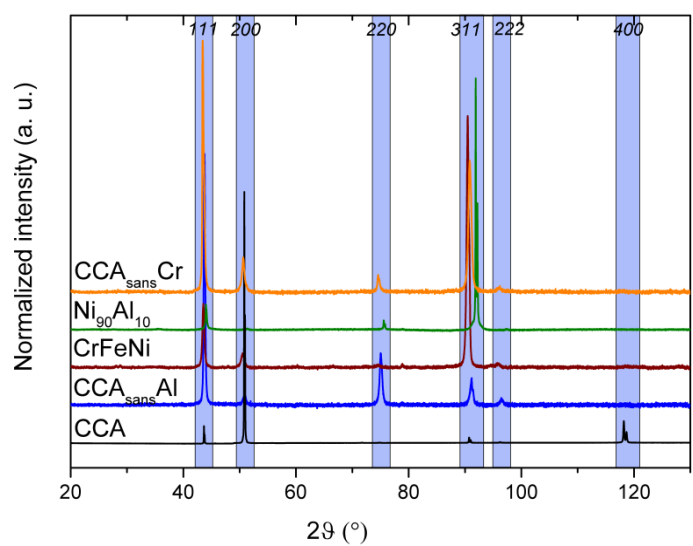

Figure 1. XRD patterns of CCA, CCA$_{sans}$Al, CrFeNi, Ni$_{90}$Al$_{10}$ and CCA$_{sans}$Cr specimens as a function of the scattering angle 2θ (20° ≤ 2θ ≤ 130°), from bottom to top, respectively. The *hkl* indices of all reflections are also shown, according to the $Fm\overline{3}m$ space group. The patterns are normalized to the highest peak intensity after background subtraction, and stacked vertically for clarity. Peak splitting is due to the Cu Kα1/Kα2 doublet.

Neither superstructure reflections in the low-*2θ* region *(20° ≤ 2θ ≤ 40°)* nor secondary phases were observed. Therefore, one can conclude that no cation ordering appears, and that high phase purity was achieved for all the specimens investigated. All XRD peaks were indexed using a simple *fcc* structure ($Fm\overline{3}m$) and lattice parameters refined accordingly. Differences in reflection intensities are given by preferred crystallite orientation. Le Bail refinements were performed,[39,40] and resulted in lattice parameters close to each other, ranging from a minimum of 3.553(1) Å (Ni$_{90}$Al$_{10}$) to a maximum of 3.601(1) Å (CrFeNi). CCA, $CCA_{sans}Al$ and $CCA_{sans}Cr$ lattice parameters were refined as 3.592(1) Å, 3.576(1) Å and 3.583(1) Å, respectively.

## 2. NEXAFS $L_{2,3}$ Spectra
### 2.1 General Analysis

The normalized NEXAFS $L_{2,3}$ spectra of Cr, Fe, Co, Ni and Cu in the $Al_8Co_{17}Cr_{17}Cu_8Fe_{17}Ni_{33}$ alloy (CCA), in the precursors ($CCA_{sans}Al$, CrFeNi, $Ni_{90}Al_{10}$) and in the corresponding pure metals are presented in Figure 2, 3, 4. For the sake of convenience, $L_{2,3}$ NEXAFS spectra are further denoted as *X(A)*, where *X* is an absorbing element in an *A* alloy, e. g. Ni(CCA) stands for a Ni $L_{2,3}$ spectrum of the CCA. The $L_{2,3}$ spectra of the alloys were recorded simultaneously in surface-sensitive TEY and in bulk-sensitive FY modes to understand whether the observed tendencies of the electronic structure transformation are the same at the surface range and in the bulk (Figure 2, 3). However, due to the self-absorption (SA) effect revealing itself when the FY is used for "thick" and concentrated (high concentration of the absorbing element) samples, only the edge position of the FY spectra obtained in the present work was analyzed, as SA causes damping of the $L_{2,3}$ spectra features and intensity redistribution above the edge.[41,42]

The $L_{2,3}$ absorption spectra of the TMs comprise of two main maxima corresponding to the $L_3$ ($2p_{3/2}\rightarrow3d$) and $L_2$ ($2p_{1/2}\rightarrow3d$) electron transitions, which are possible due to the dipole selection rules. The edge splitting is caused by spin-orbit interaction and subsequent separation of the initial atomic 2p core-level by its angular momentum into $2p_{3/2}$ and $2p_{1/2}$ states.

Besides the main $L_3$ and $L_2$ features, the spectra of Ni (Figure 2d) and Cu (Figure 3) contain satellite peaks. The satellite at about 6 eV above the Ni $L_{2,3}$ absorption edges is assigned to the final state effect caused by the admixture of a minor electronic configuration of Ni atoms to the main atomic state resulting in an energetically separate final state of the excited atom.[28,43,44] Such a satellite structure is typical for pure Ni and Ni-based metallic alloys.[25,28] The $L_3$ absorption spectrum of pure Cu (Figure 3) is characterized by a three peak pattern (features *a, b, c*) which is typical for Cu $L_3$ NEXAFS and electron energy-loss spectroscopy (EELS) spectra of pure *fcc* Cu.[45-47] Although formally 3d states of Cu are completely filled, owing to hybridization of 4s and 3d states, the 3d band of metallic Cu appears to be not fully occupied and the electron transition from 2p to empty 3d states is possible (*a* and *a'* features). In turn, the *b* and *c* features are explained by the transition of Cu 2p electrons to empty Cu 3d states with an admixture of empty Cu 4s states.[47]

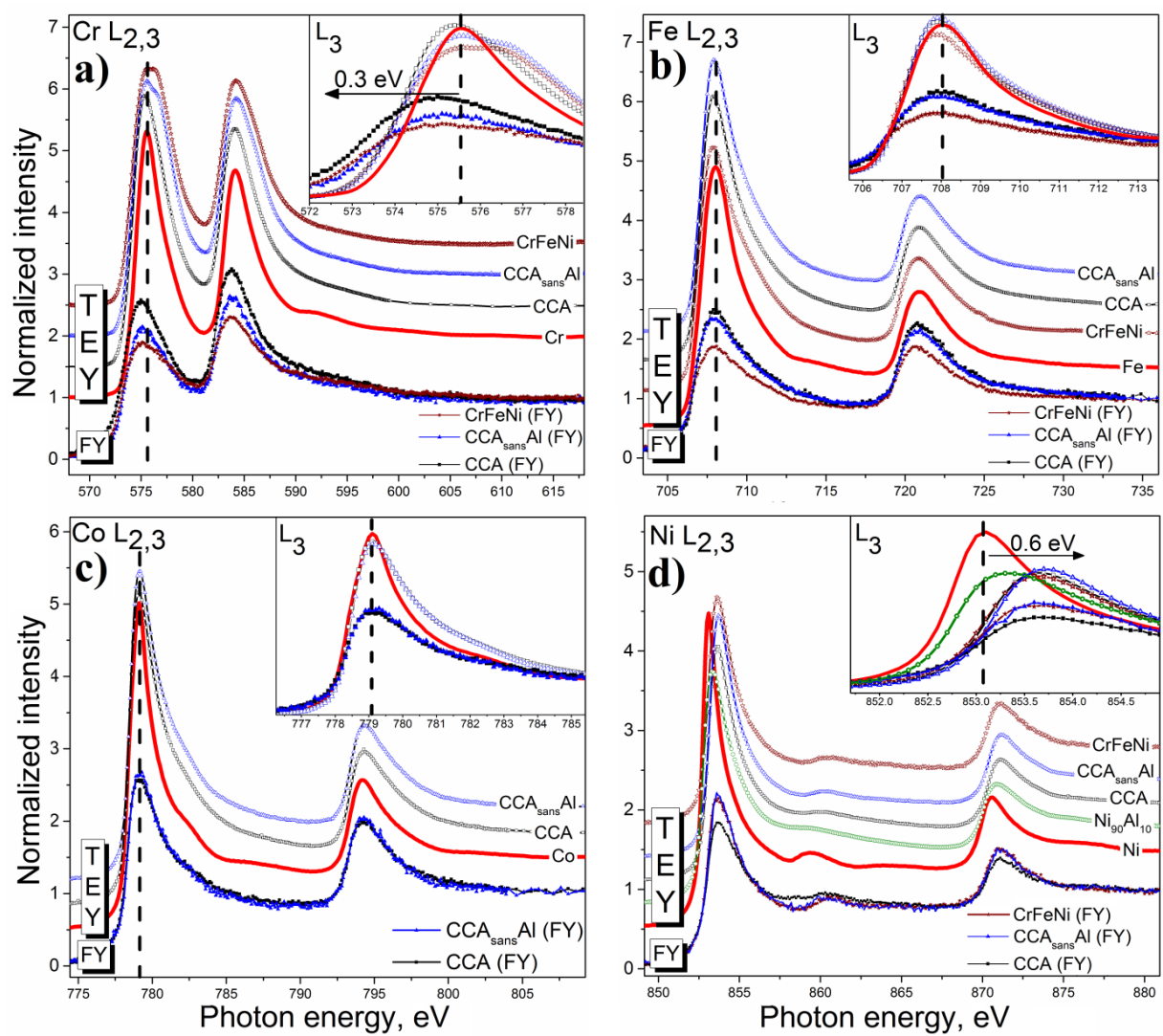

Figure 2. Normalized Cr (a), Fe (b), Co (c) and Ni (d) NEXAFS $L_{2,3}$ spectra of pure elements (TEY mode, red solid line) with a dashed line indicating the $L_3$ intensity maximum, the CCA and its precursors ($CCA_{sans}Al$, CrFeNi, $Ni_{90}Al_{10}$) recorded in TEY (empty symbols) and FY (filled symbols) modes; the inset contains the $L_3$ region of the superimposed spectra. The shift of the Cr and Ni CCA spectra compared to the pure elements is highlighted with an arrow and the corresponding value.

As can be seen from the CCA NEXAFS spectra in comparison with those of pure metals (Figure 2b and Figure 2c), small differences are observed in the Fe and Co spectra, as upon alloying only broadening of the main peaks and slight variation of the maximum intensity occurs.

Noteworthy, comparison of the Fe and Co spectra of the CCA acquired in TEY and FY modes demonstrates good agreement between the bulk (FY mode) and surface (TEY mode) signal as the edges positions remain the same. Here and in the following, a spectrum edge position is defined as the inflection point of the edge.[48,49] The edge positions of the NEXAFS spectra presented in this work are given in the SI (Table S1).

In contrast to the Fe and Co spectra, the CCA spectra of Cr, Ni and Cu deserve more attention.

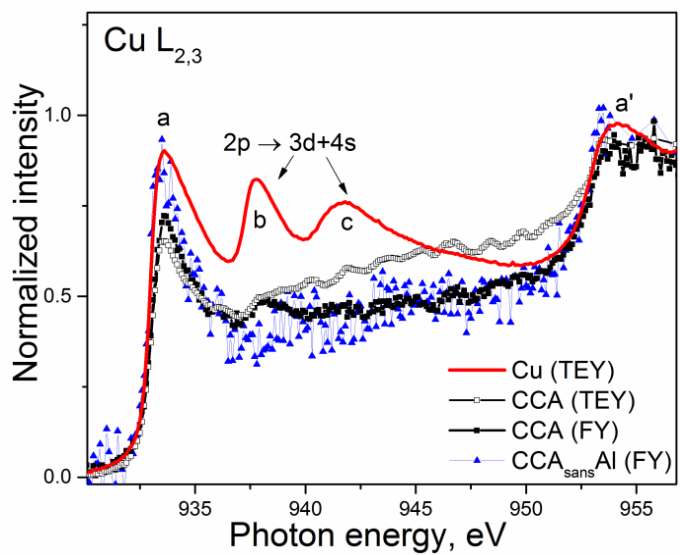

**Figure 3.** Normalized Cu NEXAFS $L_{2,3}$ spectra of pure Cu (TEY mode), the CCA (TEY and FY modes) and CCA$_{sans}$Al (FY mode). The main features of the pure Cu spectrum are denoted as a and a' (2p→3d electron transition), and b and c (2p→3d+4s transition, where 3d+4s stands for 3d states with 4s states admixture).

Figure 3 contains a pure Cu spectrum recorded in TEY mode and a Cu(CCA) recorded in both TEY and FY modes. Comparison of these spectra (Figure 3) reveals strong attenuation of the satellite structure (features *b, c*) upon CCA formation, implying an increase of Cu 4s states occupancy and/or a decrease of Cu 3d/4s states' hybridization strength. Although the TEY Cu(CCA) spectrum appeared to have a very low signal/noise ratio compared to that of the FY Cu(CCA), the TEY and FY spectra show similar general differences from the Cu pure spectrum.

Comparative analysis of the TEY Cr(CCA) and pure Cr (Figure 2a, TEY mode) spectra reveals a shift of the Cr edge position by 0.3 eV to lower energy and noticeable broadening of the $L_3$ and $L_2$ peaks of the CCA spectrum. Interestingly, the edge shift of the Cr(CCA) recorded in FY mode appears to be considerably larger (1.3 eV). The difference in the edge position of the FY and TEY spectra is ascribed to the SA effect, as the Cr L-edge spectrum is the most affected one by SA in the CCA due to the largest contribution of Cr absorption (ca. 31%, calculated at the Cr $L_3$ edge) to the total absorption compared to all other alloying elements (from ca. 13% at Fe $L_3$ edge, down to ca. 7% Cu $L_3$ edge, see the Supplementary Information (SI) for more details).[41,42]

The edge positions of the Ni(CCA) measured in TEY and FY modes shift to higher excitation energy by 0.6 eV along with decrease of the main features maximum intensity and redistribution of intensity between the edges in comparison with pure Ni. In addition, a change in energy difference between the Ni $L_3$ peak and the satellite peak is observed. Such noticeable spectrum transformation indicates particularity of chemical surroundings of Ni atoms in the CCA.

Among the possible interactions of Ni atoms with other CCA elements, the Ni-Al interaction seems to be the most expected and strongest one due to different reasons: i) it is known that Al and Ni tend to form covalent bonds in binary alloys leading to charge redistribution between Al s-, p- and Ni d-orbitals;[18-23,25] ii) Al and Ni have the highest difference in electronegativity among the elements alloyed: 1.61 (Al), 1.66 (Cr), 1.83 (Fe), 1.88 (Co), 1.91 (Ni),1.90 (Cu), according to Pauling scale. Due to the chemical interaction between Al and Ni in an Al-Ni system, the Ni 3d band transforms in a way that the maximum of the occupied d states shifts away from $E_f$ with increase of the Al concentration.[17] Such transformation can be also accompanied by energy redistribution of Ni 3d empty states and shift of empty Ni 3d states maximum away from $E_f$ with formation of a hybridization 'pseudo-gap' at $E_f$, which is typical for covalent directional bonding in metallic alloys.[17,19-21,28]

Thus, a similar interaction of Al with Ni atoms in the CCA can be expected. In order to investigate the Ni – Al interaction in details, several precursors were prepared: $Ni_{90}Al_{10}$ and $CCA_{sans}Al$. The experimental $Ni(Ni_{90}Al_{10})$ spectrum in Figure 2d supports the assumption about the Ni-Al interaction as the comparison of the $Ni(Ni_{90}Al_{10})$ and pure Ni spectra reveals transformation similar to that of the Ni(CCA): the $Ni(Ni_{90}Al_{10})$ shifts to higher energy by 0.3 eV with decreasing of the $L_3$ maximum intensity and broadening of the $L_3$ and $L_2$ peaks. Nonetheless, it was found that the $Ni(CCA_{sans}Al)$ spectrum exhibits the same transformation as for the CCA. This fact disproves the tentative assumption about the significant role of Al in the transformation of the Ni $L_{2,3}$ spectrum and, hence, the local electronic structure around Ni. Moreover, the rest of the $CCA_{sans}Al$ TMs spectra (Figure 2, 3) have no difference with the corresponding CCA spectra, demonstrating insignificant Al influence on the 3d bands structure of the TMs upon CCA formation.

Another candidate to cause the aforementioned modifications in the Ni(CCA) spectra (Figure 2d) is Cr. The difference in electronegativity between Ni and Cr is very similar to that of Ni and Al, and the influence of Cr on the Ni 3d valence states in binary alloys was previously reported.[17] Moreover, in the present work a Cr spectrum shift in the CCA and $CCA_{sans}Al$ (Figure 2a) is observed. Therefore, the Ni–Cr interaction is supposed to be strong. In order to prove this assumption, an additional precursor alloy, namely CrFeNi, was investigated. Analysis of the Ni(CrFeNi) and Cr(CrFeNi) spectra reveals spectra transformation analogous to the observed transformation of the corresponding CCA spectra: the Ni(CrFeNi) spectrum (Figure 2d) shifts by 0.6 eV to higher energy, whereas the maximum intensity of the $L_3$ peak decreases, and the Cr(CrFeNi) spectrum (Figure 2a) shifts to lower energy by 0.3 eV along with broadening of the main features. These observations support the concept that the Ni–Cr interaction is the main reason for the local electronic structure modification of Ni atoms.

In order to obtain more detailed information on the observed transformation of the local electronic structure of the TMs upon CCA formation, an integral intensity and branching ratio analysis of the NEXAFS spectra was carried out.

### 2.2 Analysis of the $L_{2,3}$ Spectra Integral Intensity

NEXAFS originates from transitions of electrons from a core level to empty electron states and the resulting NEXAFS spectrum is proportional to the X-ray absorption cross-section of the excitation process. According to the Fermi's golden rule, the cross-section is proportional to the density of empty states and the dipole matrix element squared or, in other terms, spectral density of the oscillator strengths. The integrated value of the latter equals the total number of electrons in the system.[21,50,51,52] Therefore, the changes in integral intensity of the $L_{2,3}$ spectra upon alloying reflect the changes in the occupancy of the 3d bands.

In order to extract the information about the occupancy change, an integral intensity analysis was carried out. The contribution of the resonance part (2p → 3d) to the total intensity of the NEXAFS $L_{2,3}$ spectrum was separated from the 2p → *continuum* absorption by subtraction of the two step-like functions representing the electron transition from $2p_{3/2}$ and $2p_{1/2}$ core electron states to continuum.[48,49,53] For more details about the intensity analysis and its detailed explanation, the reader is referred to the SI.

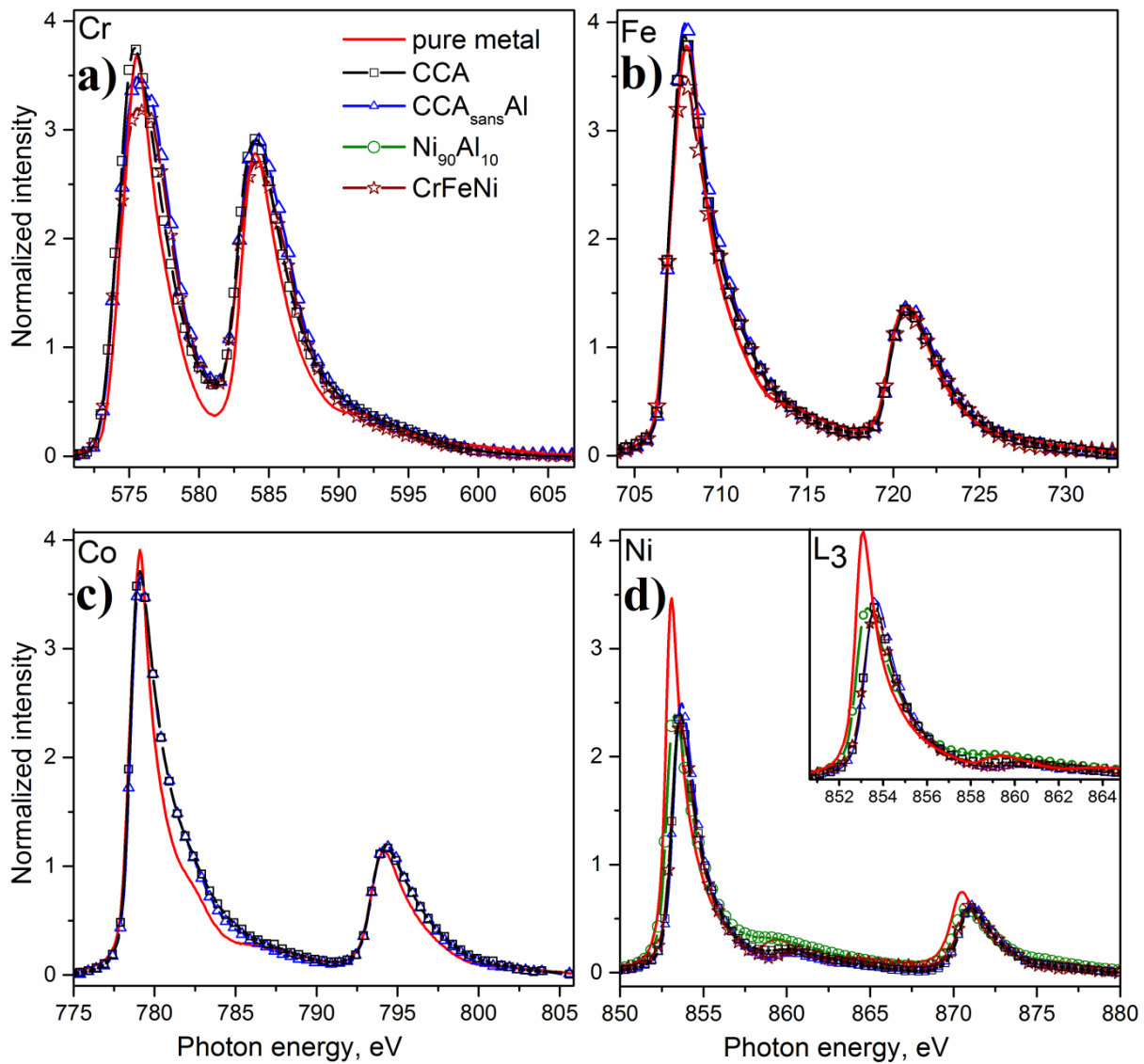

**Figure 4.** The resonance part of the NEXAFS $L_{2,3}$ spectra (TEY mode) of the CCA, precursor alloys (CCA$_{sans}$Al, CrFeNi, Ni$_{90}$Al$_{10}$) and the corresponding pure elements at the $L_{2,3}$ edges of Cr (a), Fe (b), Co (c) and Ni (d). The inset contains the $L_3$ region of the superimposed Ni $L_{2,3}$-edge spectra.

Figure 4 contains the resonance part of the $L_{2,3}$ spectra of the TM in the CCA, pure elements and reference alloys (CCA$_{sans}$Al, Ni$_{90}$Al$_{10}$, CrFeNi) recorded in TEY mode. In Figure 5, the derived total integral intensity ($I_t$) of the $L_{2,3}$ spectra is depicted as a function of the formal number of 3d electrons of the corresponding elements. The error bars were calculated taking into account the uncertainty given by the normalization procedure (cf. SI for numerical details:Table S1, S2, S3). Figure 5 shows how the $I_t$ dependence for the pure metals can be approximated by a linear function. The general trend of the $I_t$ for the pure metals is in agreement with the previous results obtained by Pearson et al., which demonstrate a nearly linear behavior of the $I_t$ for pure 3d TMs with d-shell occupancy ≥ 5 using EELS.[54] The maximum and minimum $I_t$ corresponds to the pure Cr (Figure 4a) and pure Cu (Figure 3) spectra, respectively: Cr has indeed the lowest formal occupancy of the 3d band, while Cu has the highest one among the studied elements (Figure 5, red points). The non-zero $I_t$ for the pure Cu 3d band, formally being fully occupied, demonstrates the discrepancy between the formal 3d electron number of Cu and the real occupancy of the Cu 3d states.

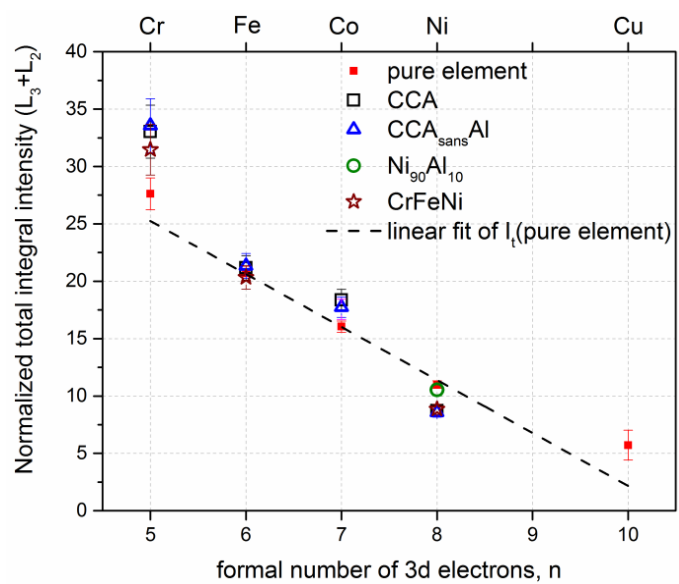

**Figure 5.** Total integral intensity of the $L_{2,3}$ NEXAFS spectra resonance part obtained from the CCA, CCA$_{sans}$Al, CrFeNi, Ni$_{90}$Al$_{10}$ and corresponding pure elements. The result of a linear approximation of the $I_t$ dependence for the pure metals is shown with a dashed line.

An $I_t$ loss of the Ni(CCA) spectrum and an $I_t$ gain of the Cr(CCA) and Co(CCA) spectra, compared with the spectra of the corresponding pure metals, are observed. On the other hand, there is negligible difference in $I_t$ of the Fe(CCA) and pure Fe spectra. These observations imply filling of the Ni 3d band along with loss of electrons by 3d bands of Cr and Co upon alloying. The same conclusion can be derived for the CCA$_{sans}$Al and CrFeNi spectra (Figure 4, Figure 5).

Meanwhile, despite the considerable attenuation of the Ni(Ni$_{90}$Al$_{10}$) L$_3$ peak compared to pure Ni L$_3$ peak, the I$_t$ values of Ni(Ni$_{90}$Al$_{10}$) and pure Ni are equal, indicating negligible variation of the Ni 3d band occupancy upon alloying Ni with Al.

### 2.3 Branching Ratio Analysis

Additional information on 3d states can be obtained through analysis of the branching ratio (BR) of the L$_{2,3}$ spectrum: BR = I(L$_x$)/(I(L$_2$)+I(L$_3$)), where I(L$_x$) is the integrated intensity of the resonance part of L$_x$ (x = 2, 3) edge. The BR parameter is known to be very sensitive to the ground state symmetry.[55,56] Many different experimental studies have shown that the BR in TM compounds[45,57-60] and pure metals[45,51,61,62] strongly deviates from the statistical ratio of 2:3 that can be obtained for a free atom in the absence of electrostatic core-valence hole interactions in the final state 2p$^5$ 3d$^{n+1}$ (n: d-electrons number in the atom initial state) of the dipole L$_{2,3}$ transition and spin-orbit coupling in the initial 3d$^n$ state, that is, in *jj* coupling limit.[55] A non-statistical BR of the NEXAFS L$_{2,3}$ spectra of TMs is a result of interplay between F$^2_{pd}$, G$^1_{pd}$, G$^3_{pd}$ Slater integrals (2p – d electrostatic interaction) in the final state, spin-orbit coupling ξ$_{2p}$ of the core-hole and spin-orbit coupling ξ$_d$ in the initial state. The BR is called statistical and equals 2/3 if the Slater integrals and ξ$_d$ are zero.

The BR values derived in this work (Figure 6, Table S3) from the pure metals Cr, Fe, Co and Ni L$_{2,3}$ spectra coincide well with the experimental BR dependence from literature.[62] The gradual growth of the BR observed with increase of 3d band occupancy for the end of the TMs series (d$^5$ – d$^9$) is rationalized by higher contribution of ξ$_{3d}$ term to the deviation of the BR from the statistical value with increase of the 3d electrons number.[62] The BR obtained for the Cr, Fe and Ni spectra of the CCA, CCA$_{sans}$Al, CrFeNi and Ni$_{90}$Al$_{10}$ exhibit very consistent discrepancy with the BR of the corresponding pure elements: the BR of Cr, Fe and Ni of the alloys spectra are larger than those of the pure metals spectra. Noteworthy, the BR of the Co spectra is the same irrespective to the sample. The BR of the Fe(CrFeNi) remains almost unchanged while the BR growth of the Cr(CrFeNi) is even larger than for the Cr(CCA) and Cr(CCA$_{sans}$Al), compared to the BR of the pure Cr spectrum.

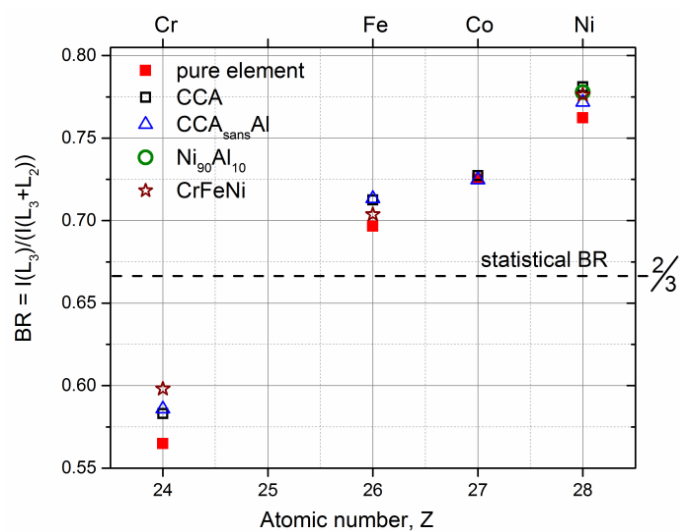

**Figure 6.** Branching ratio (BR) of the NEXAFS L$_{2,3}$ spectra recorded from the CCA, CCA$_{sans}$Al, CrFeNi, Ni$_{90}$Al$_{10}$ and corresponding pure elements. The statistical BR value is indicated by a dashed line.

The deviation of the BR of the alloys spectra from those corresponding to metal spectra can be explained by a decrease in the screening of the 2p – 3d electrostatic interaction or/and by increased spin-orbit interaction of the 3d states. Previous studies showed that increase of 3d electrons spatial localization causes growth of the BR as the screening of the 2p – 3d interaction becomes weaker.[63-66] This phenomenon was observed for a wide set of 3d TMs in thin layers with different thickness[63,65,66] and clusters of different size.[64] In both systems, increase of the layer thickness or cluster radius leads to gradual decrease in the BR.

Thus, the observed growth of the BR for the Cr, Fe and Ni spectra can be justified by higher localization of the Cr, Fe and Ni 3d states in the CCA, CCA$_{sans}$Al, CrFeNi and Ni$_{90}$Al$_{10}$ compared to the corresponding pure metals indicating introduction of covalent character into chemical bonding involving Cr, Fe and Ni upon alloy formation. The stronger change of the Cr(CrFeNi) BR compared to the CCA and CCA$_{sans}$Al is attributed to the more pronounced covalent character of the Ni–Cr bonding in CrFeNi.

### 3. 2p Core-level Photoemission Spectra

The observed energy displacement of the NEXAFS spectra can be induced either by energy shift of the initial 2p level or/and by energy redistribution of the final empty 3d states.[28] In turn, depending on the nature of the shift, the observation can point to involvement of different processes: evolution of the valence state of the absorbing atom, charge transfer between elements, change in chemical bonding character between the absorbing element and surrounding atoms.[67] In order to assign the nature of the shifts, the analysis of the 2p core levels was carried out using XPS.

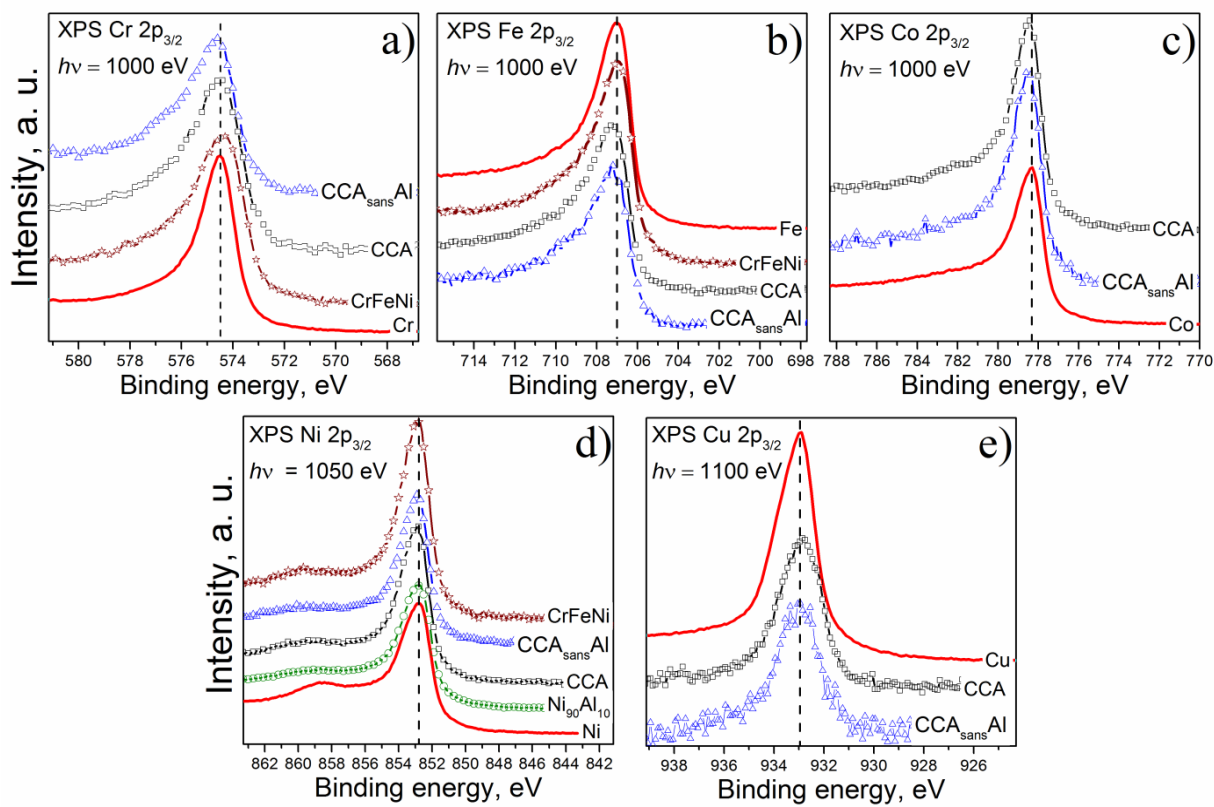

**Figure 7.** 2p XPS spectra of Cr (a), Fe (b), Co (c), Ni (d) and Cu (e) obtained in the CCA, CCA$_{sans}$Al, CrFeNi, Ni$_{90}$Al$_{10}$ and corresponding pure metals; $h\nu$ denotes the incident photon energy. The 2p$_{3/2}$ peak position of a pure element spectrum is highlighted with a dashed line.

Figure 7 displays XPS spectra of 2p$_{3/2}$ core levels of Cr, Fe, Co, Ni and Cu, while in Figure 8 the energy shifts of the 2p$_{3/2}$ XPS spectra are shown along with the NEXAFS spectra shifts (See also Table S1). An energy shift of an XPS core-level spectrum (a chemical shift) is a function of the chemical state of a probed element (change of atomic charge) and crystal potential modification.[67] As no significant changes in the XPS spectra position occur upon formation of the CCA, negligible net charge transfer between the elements and preservation of local charge neutrality around the probed elements are concluded. Moreover, the absence of the Ni and Cr 2p core lines shift in the alloys allows us to assign the nature of the Ni and Cr NEXAFS spectra shifts to the redistribution of the Ni and Cr density of the empty 3d states upon formation of the alloys.

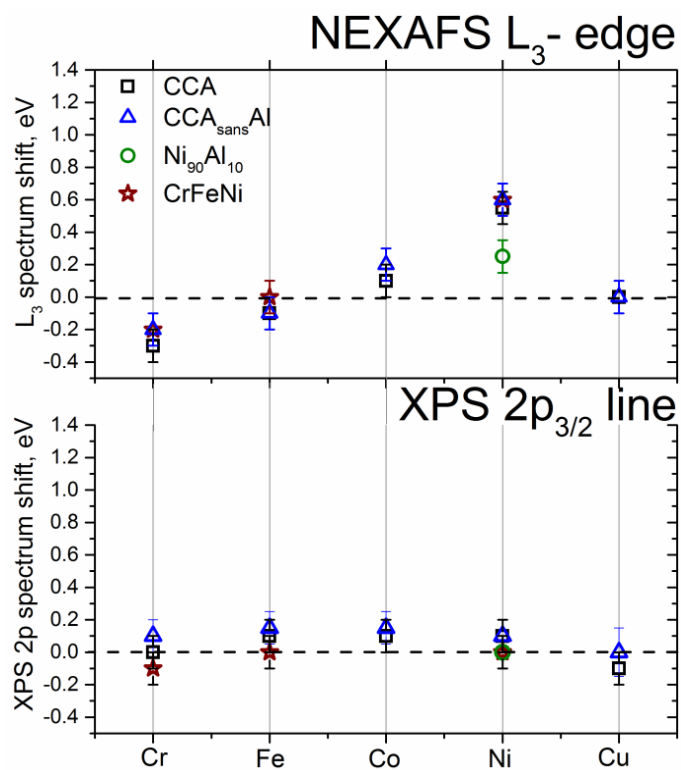

**Figure 8.** Energy shifts of 2p XPS and NEXAFS $L_3$ spectra of the CCA and its precursors (CCA$_{sans}$Al, CrFeNi, Ni$_{90}$Al$_{10}$) compared to the corresponding pure metal spectra.

### 4. Valence Band Photoemission Spectra; Cr Influence on the Ni 3d Density of States

In order to shed light on the behavior of the occupied 3d states upon CCA formation, XPS measurements of the valence bands were performed for the CCA and CCA$_{sans}$Al. Besides, a VB spectrum was recorded from an additional specimen - CCA$_{sans}$Cr (Al$_{9.6}$Co$_{20.4}$Cu$_{9.6}$Fe$_{20.4}$Ni$_{40}$) - in order to assess the influence of Cr on the occupied Ni 3d states. The VB spectra are presented in Figure 9a. The choice of the excitation photon energy (700 eV) is the result of compromise between the high signal intensity and desirable depth of analysis.

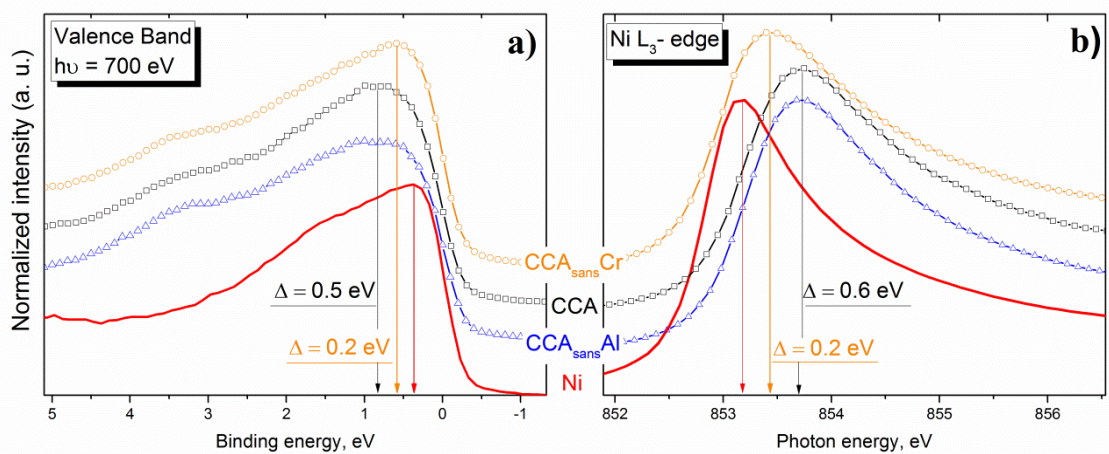

**Figure 9.** X-ray photoelectron spectra of the valence band (a) and NEXAFS Ni L$_3$-edge spectra (b) of Ni, the CCA, CCA$_{sans}$Al and CCA$_{sans}$Cr. Intensity maximum positions of the Ni, CCA and CCA$_{sans}$Cr spectra are shown by vertical arrows. The distance between the intensity maxima position of the alloys spectra and the Ni spectrum is denoted as Δ.

An XPS VB spectrum reflects distribution of total valence states density below E$_f$. However, in the case of the studied alloys, the VB signal is dominated by the Ni 3d-shell signal due to its large photoionization cross-section ($\sigma_i$) [68] and high concentration of Ni in the alloys. Noteworthy, the Cu 3d states also have large contribution to the signal despite the low Cu concentration in the alloys. Figure S1a shows VB spectra of pure metals obtained in the present work (the presented VB spectra are further denoted as VB(X), where X is the studied specimen). The mentioned domination of the Ni and Cu 3d states signal in the VB(CCA) spectrum can be demonstrated by weighting the pure metals VB spectra in respect to the 3d-shell $\sigma_i$[68] and the element percentage distribution in the CCA (Figure S1b). Contribution of the TM 4sp states and the Al sp states to the spectra can be neglected due to the low $\sigma_i$ values compared to those of the TM 3d states.[68] Thus, the observed VB spectra of the alloys mainly represent the Ni 3d occupied states, with the Cu 3d occupied states admixture appearing in a more distant energy region from E$_f$ (2 – 5 eV).

Comparison of VB(CCA), VB(CCA$_{sans}$Al) and VB(CCAsansCr) with the VB(Ni) spectrum demonstrates that the maximum of the alloys VB spectra is displayed to higher binding energy relatively to VB(Ni). For the CCA, the spectrum maximum is shifted by Δ = 0.50 ± 0.05 eV. The VB(CCA$_{sans}$Al) spectrum has no significant difference from VB(CCA). Removal of Cr from the CCA composition, preserving elements concentration ratio, leads to a decreased shift (Δ = 0.20 ± 0.05 eV) between the VB (CCA$_{sans}$Cr) and VB(Ni) maxima.

To finalize the overall picture of the Ni 3d DOS redistribution upon alloys' formation, the VB spectra of the alloys were compared to the corresponding Ni L$_3$-edge spectra, including the Ni L-edge spectrum of CCA$_{sans}$Cr (Ni(CCA$_{sans}$Cr)) (Figure 9b). Such comparison allows matching the behavior of both occupied and unoccupied Ni 3d states.

First of all, the Ni(CCA$_{sans}$Cr) spectrum is shifted only by 0.20 ± 0.05 eV compared to the pure Ni spectrum. This is a relevant decrease to the Ni spectrum shift relatively to the already observed 0.60 ± 0.05 eV shift of the Ni(CCA) and Ni(CCA$_{sans}$Al) spectra (Figure 8).

This fact implies that a tendency, similar to that of the VB spectra, is observed for the NEXAFS L$_3$-edge spectra of the alloys: the Ni edge shift decreases after the Cr is removed from the CCA alloy composition. Moreover, the corresponding shift values for the VB and NEXAFS spectra are quite close to each other. The opposite direction of the shifts of the VB maximum and Ni L$_3$-edge from E$_f$ for the CCA, compared to the pure Ni, unambiguously implies formation of a valley in the Ni 3d states density distribution at E$_f$ upon alloying.

The performed comparison of the CCA and CCA$_{sans}$Cr VB and L$_3$-edge spectra allows concluding that the presence of Cr is the main reason for the Ni 3d states density redistribution upon CCA formation.

5. Discussion

The results obtained by NEXAFS analysis of the CCA and its precursors clearly show that, although Al has the lowest electronegativity and the highest nominal atomic radius among the alloyed elements in the CCA, Al influence on the 3d states of the TMs (i. e. the 3d bands occupancy and empty states density distribution) is negligible.

One can, therefore, conclude that Al–TMs interaction in the CCA does not involve any significant transformation of the TMs 3d bands upon alloying. This fact suggests two possible cases of the Al–TMs bonding: i) the TMs 3d states are involved in the Al–TMs bonding but an even distribution of Al atoms around the TMs atoms results in blurring of the Al–TM bonding influence on the 3d states; ii) the Al–TMs chemical bonding in the CCA does not involve the localized 3d states of the TMs.

Considering the fact that directional Al–TM bonding in Al binary and ternary alloys with 3d TMs is reported to occur generally through mixing of Al 3sp and TM 3d states and redistribution of electrons between the states,[18-21,23,25,27,35] and the experimental observation of the Ni 3d band transformation for Ni$_{90}$Al$_{10}$ in the present work, a similar Al–TMs interaction should happen in the CCA. However, assuming the case when the overall influence of the Al–TMs interaction on the 3d states of the TMs is evenly shared amongst the TM elements, the obtained NEXAFS spectra might have low sensitivity to the changes in the 3d bands structure due to low Al concentration.

Nevertheless, this suggestion seems to be barely acceptable due to the Al–TM chemical preference in the CCA reported previously by Fantin et al.: higher affinity of Al for heavier 3d metals in the CCA with the Al–Ni preferred pair and less preferable Al–Cr and Al–Al pairs were found in the CCA.[34]

The absence of the Al – TM bonding influence on the 3d bands can also mean that the TMs' 3d states are not involved in the Al–TM interaction in the CCA. In this case, the interaction occurs by means of Al 3sp and TMs' 4sp delocalized states mixing with subsequent formation of delocalized bonds in order to accommodate and suppress lattice distortions given by the different radii of the elements. This hypothesis is supported by the observed filling of Cu and Ni 4p states caused by the CCA formation.[34]

On the other hand, the local electronic structure of Cr and Ni undergoes relevant transformation upon alloying of the CCA. The most noticeable changes occur in the Ni 3d band, whose occupancy increases and the occupied and empty states maximum shifts away from $E_f$ by 0.5 and 0.6 eV, respectively. This transformation is also accompanied by the decrease of the Cr 3d states occupancy and the shift of the empty 3d states maximum of Cr by 0.3 eV towards $E_f$. The shifts of the Ni and Cr 3d states density maxima in connection with the change in the occupancies suggest that the gain(loss) of electrons by the Ni(Cr) 3d states is not induced by a simple parallel energy shift of the 3d band relative to $E_f$ but by transformation of the 3d DOS shape. Based on the experimental data, a scheme is proposed (Figure 10), illustrating the transformation of the Ni (Figure 10a) and Cr (Figure 10b) 3d bands upon formation of the CCA. The DOS schematic for the pure elements was depicted focusing only on the main features in accordance with the theoretical DOS calculations from literature[69] and the XPS spectra of the Ni and Cr VB measured in this work (Figure S1a), which give insight into the occupied part of the Ni 3d and Cr 3d bands.[69]

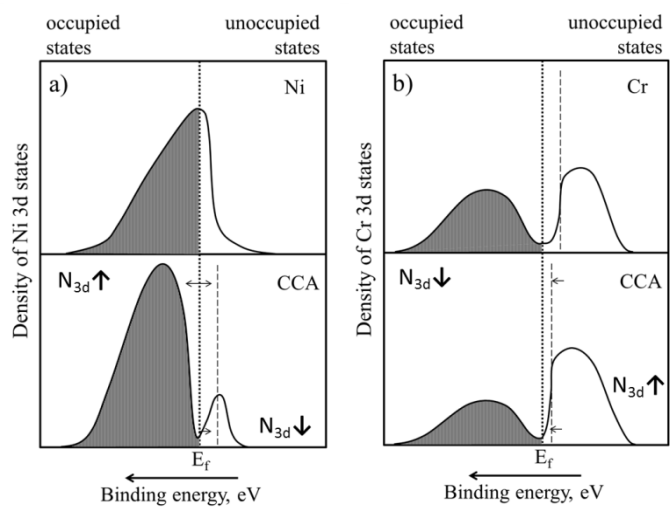

**Figure 10.** A proposed scheme for the DOS redistribution of the Ni (a) and Cr (b) 3d bands occurring upon formation of the CCA. $N_{3d}\uparrow$ and $N_{3d}\downarrow$ denote increase and decrease, respectively, in total number of states of an occupied/unoccupied part of the 3d band due to CCA formation. It is important to note that the depiction of the unchanged energy distribution of the occupied Cr 3d states in the CCA has a solely speculative character as it cannot be established from the present work.

The scheme describes the behavior of both occupied and unoccupied parts of the Ni and Cr 3d bands upon CCA formation. The information about the energy shifts of the unoccupied parts was taken from the shifts of the corresponding NEXAFS L-edge spectra. The shift of the Ni L-edge spectrum allows concluding that the maximum of the unoccupied Ni 3d states in the CCA is also shifted correspondingly from $E_f$. Similar reasoning, based on the Cr L-edge spectrum displacement, can be made for the shift of the unoccupied Cr 3d states maximum towards $E_f$.

The occupied Ni 3d states redistribution is based on the conclusion from the analysis of XPS VB spectrum of the CCA. It was found that the maximum of the occupied Ni 3d states shifts away from $E_f$ upon CCA formation.

The established tendencies in the Ni 3d states DOS unambiguously imply decrease in the Ni 3d states density at $E_f$ due to the CCA alloying, which is shown in the proposed scheme. Noteworthy, similar behavior of Ni 3d states was previously found and theoretically predicted in Ni-based binary alloys with 3d TMs having higher electronegativity compared to Ni (Sc, Ti, Cr, Mn).[17,24,28,29]

According to the scheme, hybridization of the Ni 3d states with the partner 3d element band leads to the displacement of the occupied and unoccupied Ni 3d states maximum from $E_f$ and formation of a hybridization 'pseudo-gap' at $E_f$, typical for covalent bonding character implying directionality of the bond.[17,19,26,28,29,32]

It is important to note that the Cr 3d states energy distribution of the occupied part of the 3d band in the CCA cannot be established from the present work and the unchanged distribution compared to the pure Cr, as depicted in the scheme, is speculative.

The performed analysis revealed that Cr and Ni are the most affected elements, in terms of 3d states structure transformation by the alloy formation of the CCA. The preference of Cr to bind with smaller TM elements, amongst them Ni, is corroborated by the findings of Fantin et al. on the same alloy by means of EXAFS analysis.[34]

Besides, the disappearance of the satellite structure in the Cu 3d spectrum (Figure 3, features *b, b'*) reflects filling of the Cu 4s states hybridized with the Cu 3d states and possible weakening of the hybridization strength between the Cu 4s and 3d states. Finally, Fe and Co do not exhibit any considerable density redistribution, although occupancy of the Co 3d states slightly decreases.

Interestingly, despite the 3d occupancy variation of TMs, charge neutrality of TMs atoms is preserved. This can be rationalized by previously offered concepts for TM alloys: intra-atomic redistribution of electrons (e.g. between Ni 3d and Ni 4p states)[25,28,29] and a more complex charge redistribution between elements through delocalized 4p and 4s states with zero net-charge flow.[21,23-25,28]

Additional reference alloys were studied and essential similarities in empty Cr, Fe and Ni 3d states distribution of CrFeNi, the CCA and CCA$_{sans}$Al were discovered. Furthermore, the variation of Cr, Fe and Ni 3d states occupancy and the BR value for the Cr, Fe and Ni L$_{2,3}$ spectra appeared to be very similar for these alloys. This similarity is ascribed mainly to the presence of preferable chemical interaction between Cr and Ni atoms, which is based on their electronegativity difference. Besides, it was established that removal of Cr from the CCA composition (CCA$_{sans}$Cr) leads to significant narrowing of the pseudo-gap in the Ni 3d band compared to the CCA: the shifts of the occupied and unoccupied 3d states maximum from $E_f$ become reduced to 0.20 eV. Therefore, the observed changes in the Ni 3d DOS upon CCA formation mainly result from the Ni-Cr interaction accompanied by hybridization of the Ni and Cr 3d states. Given the Ni 3d DOS transformation including the hybridization pseudo-gap appearance (Figure 10), formation of directional Ni – Cr bonds with covalent character is suggested.

The presence of such bonding can be decisive for tailoring of the material deformation properties as the bonds with uneven distribution of bonding charge density introduces anisotropy into the material matrix resulting in shear modulus increase, change in the grain boundary cohesive strength with variation in the material ductility and micro-hardness.[19-22,26,32] Moreover, the appearance of the 'pseudo-gap' in the Ni 3d band due to the hybridization of Cr and Ni 3d states can affect the phase stability, heat formation and defect propagation.[19,20,26]

Analysis of the BR parameter of the NEXAFS spectra points to weakening of the 2p – 3d electrostatic interaction screening around Cr, Fe and Ni atoms, implying higher spatial localization of 3d states around Cr, Fe and Ni atomic sites, which, in turn, supports the assumption on the covalent character introduction to the bonding between Cr and Ni atoms. Interestingly, the increase in the Fe 3d states localization occurs without any other noticeable changes in the 3d states density distribution. One should also mention that the BR does not change for Co atoms upon alloying, although Co 3d occupancy decreases. Identical behavior was discovered for the Co and Fe spectra of $CCA_{sans}Al$ and the Fe spectrum of CrFeNi. The origin of this trend remains unclear as the BR parameter is known to be quite sensitive to the symmetry, spin-state and occupancy of the 3d states.

**Conclusions**

The performed NEXAFS and XPS study of $Al_8Co_{17}Cr_{17}Cu_8Fe_{17}Ni_{33}$ allowed for the first time to experimentally investigate electronic structure and chemical interaction of the elements in a compositionally complex alloy.

It was established that the Ni 3d band gains electrons along with the redistribution of the DOS upon CCA formation: the occupied and empty Ni 3d states density maximum shifts away from $E_f$ by 0.5 and 0.6 eV, respectively, resulting in pseudo-gap formation. The Cr 3d band transformation upon alloying is opposite to that of Ni: occupancy of the band decreases and the maximum of the empty 3d states shifts slightly by 0.3 eV towards $E_f$. Besides these transformations, the Cu 4s states with admixture of Cu 3d states symmetry (3d/4s) gain electrons, the Co 3d band loses electrons and the Fe 3d band preserves the structure of the empty states and the band occupancy upon alloying.

According to the XPS analysis, the net charge transfer between the elements is negligible, despite the observed changes in the 3d bands occupancy. This fact is rationalized by compensation of the changes in the 3d states occupancy by redistribution of delocalized s- and p- electrons of the TMs in a way that the net charge flow on and off sites in the alloy is absent.

It was found that Al contribution to the 3d electronic structure changes is negligible, despite the expected Al–TMs covalent bonding.

The most noticeable changes in the TMs electronic structure are attributed to the chemical interaction of Ni and Cr atoms and formation of directional bonds with covalent character, which result in occupancy change and DOS redistribution of the Ni and Cr 3d bands. The presence of such bonding in the CCA is expected to make possible tailoring the material deformation properties by varying covalent character of the Ni-Cr bonds.

**Acknowledgement**

Allocation of beamtime by Helmholtz-Zentrum Berlin is gratefully acknowledged. The authors are grateful to the German Research Foundation (DFG) for the financial support [grant numbers: GL 181/57-1, BA 1170/391-1], through the Priority Programme SPP 2006 "Compositionally Complex Alloys - High Entropy Alloys (CCA - HEA)", and to the German-Russian Interdisciplinary Science Center (G-RISC) [grant numbers: P-2018a-8, T-2019a-4] funded by the German Federal Foreign Office via the German Academic Exchange Service (DAAD). C. Leistner and C. Förster (HZB) are acknowledged for help in sample preparation.

Supporting Information for

# Chemical Interaction and Electronic Structure in a Compositionally Complex Alloy: a Case Study by means of X-ray Absorption and X-ray Photoelectron Spectroscopy.

S. Kasatikov[1,2], A. Fantin[2,3], A.M. Manzoni[2,4], S. Sakhonenkov[1], A. Makarova[5], D. Smirnov[6], E. Filatova[1*] & G. Schumacher[2,3]

[1] Institute of Physics, St-Petersburg State University, Ulyanovskaya Str. 3, Peterhof 198504 St. Petersburg, Russia

[2] Helmholtz-Zentrum Berlin, D-14109 Berlin, Germany

[3] Technische Universität Berlin, D-10623 Berlin, Germany

[4] Bundesanstalt für Materialforschung und -prüfung, Abteilung Werkstofftechnik, Berlin, Germany

[5] Institut für Festkörper- und Materialphysik, Technische Universität Dresden, 01062 Dresden, Germany

[6] Physikalische Chemie, Institut für Chemie und Biochemie, Freie Universität Berlin, 14195 Berlin, Germany

*Corresponding author: E. O. Filatova, feo@ef14131.spb.edu

**List of Contents**

1. **Description of the procedures used for the NEXAFS spectra normalization and integral intensity analysis**

   **1.1** *Normalization of the NEXAFS spectra*

The $L_{2,3}$ NEXAFS spectra were acquired in a wide energy range: at least 50 eV (100 eV for the pure metals spectra) below the absorption edge and 60 – 100 (150 eV for the pure metals spectra) eV above the absorption edge. Normalization of the NEXAFS spectra was performed in two steps using the software ATHENA[1]: i) pre-edge background removal ii) edge step normalization using a polynomial function interpolated far above the edge region.

   **1.2** *Integral intensity analysis of the NEXAFS spectra*

Contribution of the resonance part (2p → 3d) to the total intensity of the NEXAFS $L_{2,3}$ spectrum was separated from the 2p → *continuum* absorption by subtraction of the two step-like functions representing the electron transition from $2p_{3/2}$ and $2p_{1/2}$ states to continuum:

$$\sigma(E) = \sigma_0 \left\{ \frac{1}{2} - \frac{1}{\pi} tan^{-1}((E - E_0)/(\gamma/2)) \right\} \qquad (1),$$

where $\sigma_0$ - height of the edge jump, $E_0$ – position of the absorption edge, $\gamma = 1/\tau$ – 2p atomic level width, where τ denotes the life-time of the core hole.

The edge positions ($E_0$) of the spectra were determined as the inflection point of the $L_3$ and $L_2$ edges, respectively (see Table S1). The $\gamma$ parameter was obtained from literature.[2] All the parameters used for the subtraction procedure are presented in Table S2. Larger values of $\gamma$ used for the $L_2$ continuum absorption step simulation are explained by the reduction of the $2p_{3/2}$ core-hole lifetime τ, compared to that of the $2p_{1/2}$ core-hole, induced by $L_2L_3V$ Coster-Kronig transitions.[3]

Calculated integral intensity values (total and partial) and corresponding integration limits are presented in Table S3. The separation point between the $L_3$ and $L_2$ resonance parts contribution to the total spectra was chosen at the intensity minimum between the $L_3$ and $L_2$ edges. Contribution of the $L_3$ and $L_2$ resonance parts to the overall intensity of Cr spectra was determined by means of resonance spectra decomposition due to the vicinity of the Cr $L_3$ and $L_2$ peaks. The decomposition procedure was performed using the software CasaXPS[4]. The resonance part of the Cr $L_{2,3}$ spectra was fitted with two asymmetric peaks representing the $L_3$ and $L_2$ main features with the same line-shape of the peaks (GL(30)T(0.9)) and distance between the peaks (8.6 eV) for each spectra.

The integral intensity uncertainty was given by the normalization procedure i. e. integral intensity scatter induced by the freedom of the pre-edge background removal and edge jump normalization.

**Table S1: Energy positions of the NEXAFS spectra L-edges and 2p$_{3/2}$ core XPS lines measured in the present work.**

| Sample name | Energy position, eV; measurement uncertainty: ± 0.1 eV | | | | | | | | | | | | | | |
|---|---|---|---|---|---|---|---|---|---|---|---|---|---|---|---|
| | Cr | | | Fe | | | Co | | | Ni | | | Cu | | |
| | 2p$_{3/2}$ | L$_3$ | L$_2$ | 2p$_{3/2}$ | L$_3$ | L$_2$ | 2p$_{3/2}$ | L$_3$ | L$_2$ | 2p$_{3/2}$ | L$_3$ | L$_2$ | 2p$_{3/2}$ | L$_3$ | L$_2$ |
| pure metal | 574.5 | 574.5 | 583.8 | 707.2 | 707.2 | 720.2 | 778.3 | 778.3 | 793.3 | 852.8 | 852.8 | 870.0 | 933.0 | 933.0 | 952.9 |
| CCA | 574.5 | 574.2 | 583.5 | 707.3 | 707.1 | 720.1 | 778.4 | 778.4 | 793.4 | 852.9 | 853.4 | 870.6 | 932.9 | 933.0 | - |
| CCA$_{sans}$Al | 574.6 | 574.3 | 583.6 | 707.3 | 707.1 | 720.1 | 778.4 | 778.5 | 793.5 | 852.9 | 853.4 | 870.6 | 933.0 | 933.0 | - |
| CrFeNi | 574.4 | 574.3 | 583.6 | 707.2 | 707.2 | 720.2 | - | - | - | 852.8 | 853.4 | 870.6 | - | - | - |
| Ni$_{90}$Al$_{10}$ | - | - | - | - | - | - | - | - | - | 852.8 | 853.1 | 870.3 | - | - | - |

**Table S2. Parameters of the step-like function used for the resonance part extraction of the NEXAFS spectra obtained in the present work.**

| Sample name | Cr | | Fe | | Co | | Ni | |
|---|---|---|---|---|---|---|---|---|
| | γ(L$_3$;L$_2$) | σ$_0$(L$_3$;L$_2$) | γ(L$_3$;L$_2$) | σ$_0$(L$_3$;L$_2$) | γ(L$_3$;L$_2$) | σ$_0$(L$_3$;L$_2$) | γ(L$_3$;L$_2$) | σ$_0$(L$_3$;L$_2$) |
| pure metal | 0.32;0.76 | 0.674;0.337 | 0.41;1.14 | 0.674;0.337 | 0.47;1.13 | 0.674;0.337 | 0.53;0.98 | 0.667;0.334 |
| CCA | 0.32;0.76 | 0.674;0.337 | 0.41;1.14 | 0.67;0.335 | 0.47;1.13 | 0.674;0.337 | 0.53;0.98 | 0.67;0.335 |
| CCA$_{sans}$Al | 0.32;0.76 | 0.674;0.337 | 0.41;1.14 | 0.67;0.335 | 0.47;1.13 | 0.674;0.337 | 0.53;0.98 | 0.665;0.333 |
| CrFeNi | 0.32;0.76 | 0.668;0.334 | 0.41;1.14 | 0.67;0.335 | - | - | 0.53;0.98 | 0.67;0.335 |
| Ni$_{90}$Al$_{10}$ | - | - | - | - | - | - | 0.53;0.98 | 0.667;0.334 |

**Table S3. Total and partial integral intensity values (with the corresponding limits of integration) and BR values derived from the NEXAFS spectra obtained in the present work.**

| Element | Sample name | Total intensity | | L3 | | L2 | | BR |
|---|---|---|---|---|---|---|---|---|
| | | Value | Limits, eV | Value | Limits, eV | Value | Limits, eV | |
| Cr | Cr pure | 27.6 ± 0.5 | 565.4 - 605.4 | 14.1 ± 0.2 | decomposed | 10.9 ± 0.2 | decomposed | 0.565 ± 0.012 |
| | CCA | 33.0 ± 0.7 | 565.4 - 605.4 | 18.2 ± 0.3 | decomposed | 13.0 ± 0.3 | decomposed | 0.583 ± 0.015 |
| | CCA$_{sans}$Al | 33.6 ± 0.7 | 565.4 - 605.4 | 18.8 ± 0.3 | decomposed | 13.3 ± 0.3 | decomposed | 0.586 ± 0.015 |
| | CrFeNi | 31.5 ± 0.7 | 565.4 - 605.4 | 17.8 ± 0.3 | decomposed | 12.0 ± 0.3 | decomposed | 0.597 ± 0.016 |
| Fe | Fe pure | 20.7 ± 0.2 | 693.9 - 735.9 | 14.41 ± 0.14 | 693.9-718.0 | 6.28 ± 0.06 | 718.0-735.9 | 0.697 ± 0.012 |
| | CCA | 21.2 ± 0.4 | 693.9 - 735.9 | 15.1 ± 0.3 | 693.9-718.9 | 6.08 ± 0.12 | 718.9-735.9 | 0.71 ± 0.02 |
| | CCA$_{sans}$Al | 21.3 ± 0.4 | 693.9 - 735.9 | 15.2 ± 0.3 | 693.9-718.9 | 6.12 ± 0.12 | 718.9-735.9 | 0.71 ± 0.02 |
| | CrFeNi | 20.3 ± 0.4 | 693.9 - 735.9 | 14.3 ± 0.3 | 693.9-718.9 | 6.02 ± 0.12 | 718.9-735.9 | 0.70 ± 0.02 |
| Co | Co pure | 16.1 ± 0.2 | 769.1 - 807.1 | 11.63 ± 0.16 | 769.1-791.0 | 4.41 ± 0.06 | 791.0-807.1 | 0.725 ± 0.014 |
| | CCA | 18.4 ± 0.4 | 769.1 - 807.1 | 13.4 ± 0.3 | 769.1-791.0 | 5.0 ± 0.1 | 791.0-807.1 | 0.727 ± 0.022 |
| | CCA$_{sans}$Al | 17.7 ± 0.4 | 769.1 - 807.1 | 12.9 ± 0.3 | 769.1-791.0 | 4.9 ± 0.1 | 791.0-807.1 | 0.725 ± 0.024 |
| Ni | Ni pure | 11.0 ± 0.1 | 843.2 - 880.6 | 8.39 ± 0.08 | 843.2-867.2 | 2.62 ± 0.02 | 866.3-880.6 | 0.76 ± 0.01 |
| | CCA | 8.7 ± 0.1 | 843.1 - 881.1 | 6.81 ± 0.08 | 843.1-867.9 | 1.91 ± 0.02 | 868.8-881.1 | 0.781 ± 0.013 |
| | CCA$_{sans}$Al | 8.6 ± 0.1 | 843.1 – 881.1 | 6.62 ± 0.08 | 843.1-868.1 | 1.96 ± 0.02 | 868.1-882.0 | 0.772 ± 0.013 |
| | CrFeNi | 8.9 ± 0.1 | 843.1 – 881.1 | 6.87 ± 0.08 | 843.1-868.1 | 1.98 ± 0.02 | 868.1-882.0 | 0.776 ± 0.013 |
| | Ni$_{90}$Al$_{10}$ | 10.5 ± 0.1 | 843.1 – 883.1 | 8.19 ± 0.08 | 843.1-867.9 | 2.34 ± 0.02 | 868.8-883.1 | 0.778 ± 0.011 |

| | | | | | | | | | |
|---|---|---|---|---|---|---|---|---|---|
| Cu | Cu pure | 5.7 ± 0.5 | 926.8 – 958.8 | - | - | - | - | - |

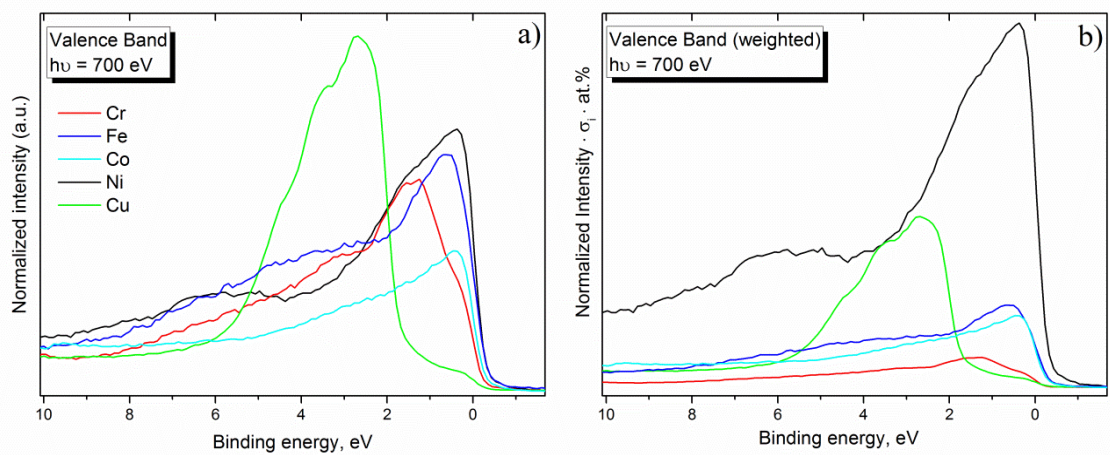

**Figure S1.** Valence band spectra of pure metals (Cr, Fe, Co, Ni and Cu) recorded at 700 eV incident photon energy (a). In order to assess the signal contribution from a 3d band of each TM element to the VB spectrum of the CCA, the recorded spectra were weighted (b) with corresponding photoionization cross-section $\sigma_i$[5] and atomic percentage of the element (at. %) in the CCA.